# Electric Fields Turn Vacancy-trapping Grain Boundaries into Li-Ion Pathways in $Li_2O$

*Youjeong Choi [a], Po-Yen Chen [a], Tasuku Sugiura [b], and Teruyasu Mizoguchi [a, b]**

[a] Department of Materials Engineering, The University of Tokyo, Tokyo 113-8656, Japan

[b] Institute of Industrial Science, The University of Tokyo, Tokyo 153-8505, Japan

* Email: teru@iis.u-tokyo.ac.jp

**ABSTRACT**

Ion transport in solids has conventionally been understood from equilibrium diffusion, yet ions in batteries are driven by electric fields. Does field-driven ion transport follow equilibrium behavior? We examined this question in bulk $Li_2O$ and at its grain boundaries (GBs) using electric-field molecular dynamics combining a machine learning interatomic potential with a Bader charge prediction model. Although Li diffusion in cubic $Li_2O$ is isotropic at zero field, an electric field renders it orientation dependent because Li hops along a limited set of crystallographic pathways that the field biases unequally. At GBs, this anisotropy becomes a qualitative difference: a GB that strongly traps Li vacancies but exhibits a strong field response releases them under a field, whereas a weakly responsive GB retains them despite weaker vacancy trapping. These results show that field-driven GB transport is governed not by static thermodynamics alone but by the angle between the electric field and the diffusion pathways of the constituent grains.

It is essential to understand ion transport behavior under an applied electric field at the atomic scale in order to design materials with high ionic conductivity and improve the performance of various electrochemical energy-storage systems.[1–3] Molecular dynamics (MD) simulations have been widely used to elucidate ion diffusion pathways and dynamical properties. Ionic conductivity is commonly estimated from diffusion coefficients obtained in equilibrium MD via the Nernst–Einstein relation,[4] which implicitly assumes that field-driven transport is a linear response to equilibrium diffusion. In practical electrochemical systems, ion transport is driven by both thermal diffusion and electrochemical driving forces, with local electric fields in electric double layers and space-charge regions reaching ~$10^7$ V/cm, where the field work over a single ion hop becomes comparable to the thermal energy and linear-response assumption may break down. [5–8]. Furthermore, grain boundaries (GBs) in ionic solids have local atomic structures distinct from the bulk and can either hinder ion migration or provide pathways for ion transport.[9,10] These GB effects have been investigated mainly by equilibrium MD, and it remains unclear whether GBs that trap or block ions at equilibrium behave in the same way under a strong electric field. Does field-driven ion transport follow equilibrium behavior, and if not, what governs it?

To gain atomic-scale insight into this long-standing question regarding field-driven ion behavior and the effects of lattice defects, we investigated Li-ion transport under an external electric field in the bulk and grain boundary structures of $Li_2O$. $Li_2O$ is a representative Li–O ionic solid and adopts a simple cubic antifluorite structure, which facilitates structural modeling and the interpretation of ion transport behavior. In addition, $Li_2O$ is one of the major inorganic components of the solid electrolyte interphase (SEI) in practical batteries.[11–13] The SEI requires efficient Li-ion transport and develops GBs during growth, which are known to induce anisotropic Li-ion diffusion.[9] First-principles MD simulations that explicitly account for external fields in complex interfacial structures, as required for the present study, are computationally demanding in terms of both time and cost. However, machine-learning interatomic potentials (MLIPs) offer a promising route to extend the accessible spatial and temporal scales while retaining accuracy.[1,14–16] By coupling an MLIP with an atomic-charge prediction model, the forces acting on atoms in response to an electric field can be described.

In this study, we combined the fine-tuned MACE potential with the GNN-based Bader charge prediction model for $Li_2O$ and applied the resulting framework to MD simulations under an external electric field. Li-ion transport in bulk and GB structures of $Li_2O$ was then analyzed and compared with and without an applied electric field. Diffusion in cubic $Li_2O$ is isotropic without an electric field, whereas under strong local electric fields comparable to those that can arise near interfaces, Li-ion transport becomes dependent on crystallographic orientation. This behavior reflects the restriction of Li diffusion pathways to the ⟨100⟩ directions and leads to qualitative differences in long-range ion transport at GBs. Under an electric field, a GB that has a high Li vacancy segregation energy but exhibits a strong field response releases Li vacancies and forms a pathway for ion transport. In the opposite case, however, the GB continues to trap the Li vacancies. These results demonstrate that field-driven ion transport at GBs is governed not solely by static thermodynamic quantities but also by the crystallographic orientations of the constituent grains. Thus, equilibrium analysis alone may not fully capture ion transport at interfaces.

The field-induced force acting on each atom is given by the product of its effective charge and the electric field. Born effective charges (BECs), which describe the polarization response to atomic displacement, provide a theoretically rigorous description and have been widely used in electric-field MD simulations of materials such as ferroelectrics.[17–19] However, because BECs are defined from the polarization response of the entire crystal, their values can fluctuate strongly in regions with distorted local structures such as GBs. In some cases, individual atoms can even exhibit anomalously large or divergent BEC values.[20,21] Using such values to calculate field-induced forces can result in nonphysical forces acting on atoms. This issue is particularly relevant to the present study because Li-ion transport is mediated by vacancies. Indeed, we observed anomalous BEC values for atoms near Li vacancies and Schottky defects in $Li_2O$. In contrast, Bader charges are static atomic charges obtained by partitioning the electron density in real space and can be defined more robustly even in locally distorted environments.[22,23] For defect-free bulk $Li_2O$, electric-field MD simulations using charge prediction models trained on defect-free structures with BECs and Bader charges as targets, respectively, showed only a 3.6% difference in the mean Li MSD, indicating similar Li-ion dynamics. These results are summarized in **Supporting Information S1**. Based on this comparison, we adopted Bader charges as the effective target charges and constructed the corresponding charge prediction model. Previous studies have also developed GNN-based models using Bader charges.[24,25] These models successfully predicted structure-dependent partial Bader charges and reproduced structural behavior. Details of the dataset construction, model development, and computational settings are provided in the **Methods** section. The fine-tuned MACE potential showed good agreement with DFT for both energies and atomic forces. The Bader charge prediction model also achieved high predictive accuracy. Quantitative validation results together with a comparison with the BEC prediction model are provided in **Supporting Information S1**.

### Part 1: Electric-field response of bulk $Li_2O$ along different crystallographic orientations

The combined models were applied to bulk $Li_2O$ to investigate the crystal orientation dependence of Li-ion transport under an external electric field. **Figure 1**a shows the cubic $Li_2O$ cell and the direction of the electric field. Ten representative low-index crystallographic orientations were selected as the field directions. Owing to cubic symmetry, all crystallographic orientations can be reduced to the fundamental triangle of the inverse pole figure with [100], [110], and [111] as its vertices. In addition to these three vertex orientations, the selected ten orientations are distributed nearly uniformly along the edges and within the interior of the triangle, thereby broadly covering the crystallographic orientation space. The composition, Li vacancy concentration, and cell length along the field direction for each bulk structure are summarized in **Table S1**, and the ten crystal structures viewed along the field direction are shown in **Figure S5**.

The diffusion coefficient is a physical property represented by a symmetric second-rank tensor $D_{ij}$. According to Neumann's principle, a physical-property tensor must remain invariant under the symmetry operations of the crystal point group.[26,27] For cubic $Li_2O$ with point group $m\bar{3}m$, the diagonal components of $D_{ij}$ are equivalent and the off-diagonal components vanish. Therefore, the diffusion coefficient of $Li_2O$ without an electric field is isotropic and independent of crystallographic orientation. To confirm this, three independent MD simulations were performed for each structure without an electric field, and the diffusion coefficients corrected for Li vacancy concentration, $D^0/c_v$, are shown in **Figure 1**b. Although slight variations were observed due to differences in cell size, the corrected diffusion coefficient averaged $1.80 \times 10^{-5}$ $cm^2/s$, with inter-structure deviations remaining within approximately 7.8 %. This indicates small differences in Li-ion diffusion among the bulk structures and is consistent with the expectation that Li-ion diffusion does not strongly depend on crystallographic orientation without an electric field.

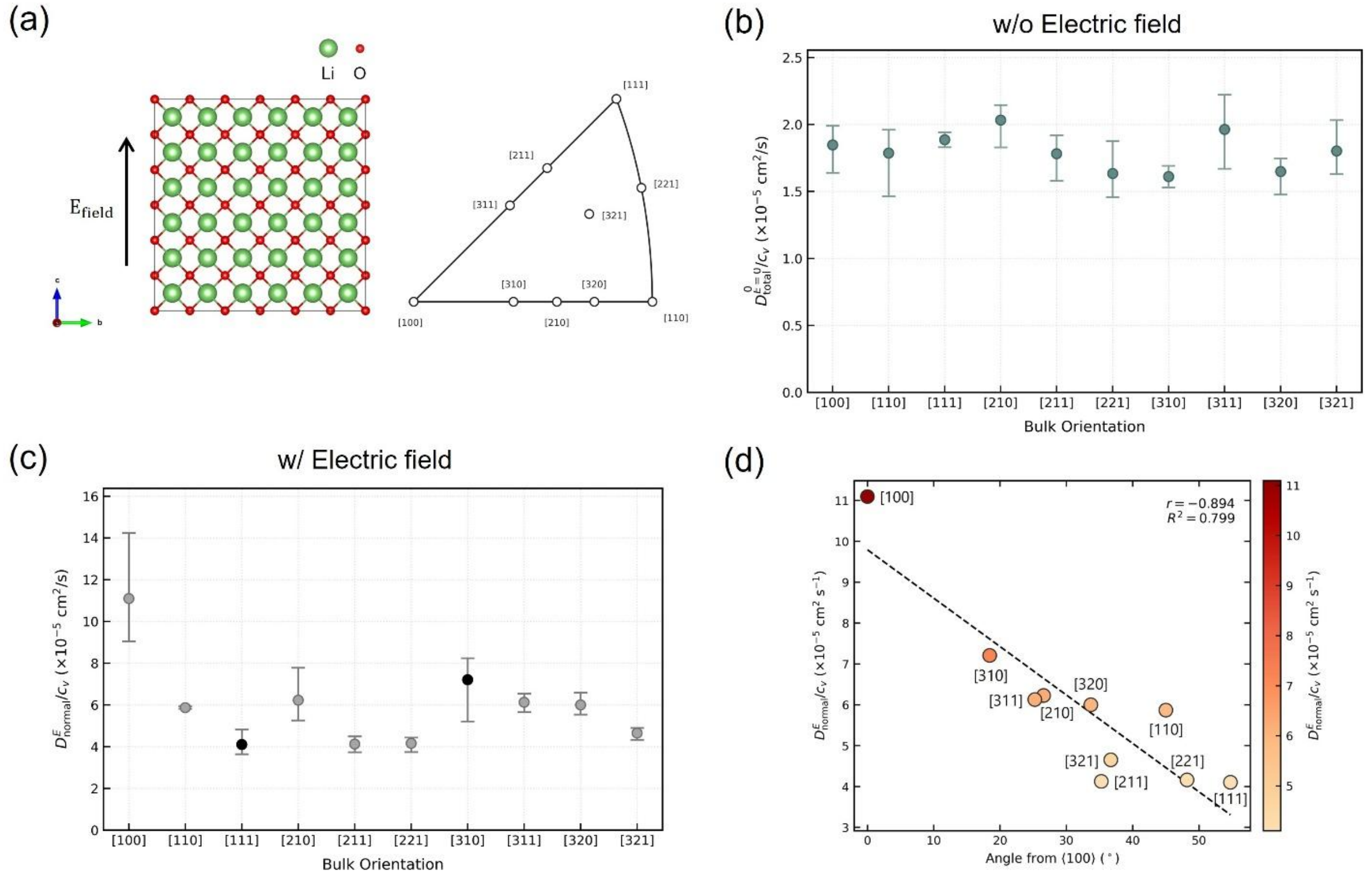

**Figure 1**. Crystallographic orientation dependence and electric-field response of Li-ion transport in bulk Li2O. (a) Cubic Li2O with the applied field direction and a cubic inverse pole figure showing the ten representative low-index orientations considered in this study. (b) Total three-dimensional Li-ion diffusion coefficients for each orientation without electric field (D0/cv). (c) Li-ion diffusion coefficients for each orientation along the field direction with electric field (DE/cv). (d) Relationship between the angle θ to the nearest ⟨100⟩ direction and DE/cv. The color indicates the magnitude of the electric-field response along the field direction. All diffusion coefficients were normalized by the Li vacancy concentration cv. The drift component induced by the electric field was removed by subtracting the average displacement of all Li ions from the displacement of each Li ion.

In contrast, **Figure 1**c shows the results obtained by applying an electric field to each structure. Three independent MD simulations were again performed, and the corrected diffusion coefficient along the applied field direction, $D^E/c_v$, was quantitatively compared among the different orientations. This value increases substantially for all crystallographic orientations under the electric field. More importantly, unlike the zero-field case, clear differences in Li-ion mobility are observed. The [100] and [310] orientations show the largest average field responses, whereas [111] orientation shows the smallest. The difference reaches approximately 35 %, far exceeding the 7.8 % statistical variation observed without an electric field. As discussed above, the zero-field diffusion coefficient is isotropic because of cubic symmetry. Therefore, such anisotropy cannot arise within the linear-response regime and instead originates from the nonlinear response induced by the strong electric field. To confirm this, the field-strength dependence was examined for the [100], [310], and [111] orientations at 0, 0.050, 0.075, 0.100 V/Å (**Figure S6**). At 0.050 V/Å, the three orientations already showed slight differences in mobility, which became more apparent with increasing field strength. It was particularly pronounced at 0.100 V/Å, while [111] consistently showed the weakest response. This field-dependent enhancement directly demonstrates that orientation dependence is nonlinear effect of the field. Cubic $Li_2O$ does not distinguish crystallographic orientations without an electric field, whereas clear orientation dependence emerges under an electric field.

This orientation dependence itself is a notable result. Without an electric field, Li-ion diffusion is isotropic but becomes strongly dependent on crystallographic orientation under an applied field. One possible explanation is that Li ions can migrate only along a limited number of crystallographic directions. In antifluorite $Li_2O$, Li ions occupy a simple cubic sublattice with a spacing of $\frac{a}{2} = 2.31$ Å and vacancy-mediated hopping is expected to occur along one of the six $\langle 100 \rangle$ directions ($[100][\bar{1}00][010][0\bar{1}0]\ldots$). Consistent with this geometry, the $\langle 100 \rangle$ exhibits the strongest response to the field. The angle $\theta$ between each crystallographic orientation and its nearest $\langle 100 \rangle$ direction follows the order 100 (0°) < 310 (18.4°) < 311 (25.2°) < 210 (26.6°) < 320 (33.7°) < 211 (35.3°) < 321 (36.7°) < 110 (45.0°) < 221 (48.2°) < 111 (54.7°). **Figure 1**d shows the relationship between this angle and the response along the applied field direction. Smaller angles tend to show stronger responses, whereas larger angles tend to show weaker responses. Among these orientations, [111] is farthest from all six hopping directions and is therefore geometrically the least favorable for Li-ion migration under an electric field.

This result is important because it suggests a limitation in estimating ionic conductivity from equilibrium MD simulations using the Nernst–Einstein relation.[4] Under the electric field relevant to actual battery devices, this approach may introduce orientation-dependent errors. The results also indicate that the orientations of the two grains forming GBs can be a primary factor governing Li-ion transport under an electric field. This effect is examined for GB structures in the following section.

**Part 2: Li-ion transport in $Li_2O$ grain boundaries**

To examine whether orientation dependence identified in **Part 1** governs Li-ion transport at GBs, [111] and [310] were selected as GB normal directions representing relatively weak and strong electric-field responses in bulk $Li_2O$. Based on these orientations, Σ3(111) and Σ5(310) symmetric tilt GBs were constructed with tilt axes of $[1\bar{1}0]$ and [001], respectively. The corresponding cells contained 432 and 480 atoms with GB normal lengths of 32.11 and 29.31 Å, providing sufficiently large GB and bulk-like regions. For these GBs, relative translation searches based on the cell of non-identical displacement (CNID) were performed using interface_master.[28,29] The mirror-symmetric GBs generated by interface_master at CNID = [0, 0, 0] were denoted as the Σ3 and Σ5 initial GBs and are shown in **Figure S7**. Starting from these initial GBs, 16 CNID candidates were generated for the Σ3 GB using a 4×4 grid that accounts for its higher symmetry. For the Σ5 GB, 64 candidates were generated using a finer 8×8 grid and their energies were compared. The overall energy distributions of the CNID candidate structures are shown in **Figure S7**.

The most stable GBs obtained from the CNID search were used for further analysis and are referred to as the Σ3(111) GB and Σ5(310) GB. Their atomic structures viewed normal to the GB plane are shown in **Figure 2**. For region-specific analysis, each structure was divided into three regions: GB, transition, and bulk-like regions. The region surrounding the GB plane was defined as the GB region, while the region retaining the cubic structure of bulk $Li_2O$ was defined as the bulk-like region. The deformed region between them was classified as the transition region. These regions were determined based on the Li vacancy formation energies shown in **Figure S8** and the local Li distortions shown in **Figure S9**. The GB energies calculated using the fine-tuned MACE potential were 0.663 $J/m^2$ for the Σ3(111) GB and 1.363 $J/m^2$ for the Σ5(310) GB. Although the Σ3(111) GB was not included in the training data, its GB energy agrees well with our DFT value of 0.659 J/m$^2$ with the previously reported value of 0.689 J/m$^2$,[9] and falls within the range of DFT-calculated GB energies reported for typical oxides.[30–32]

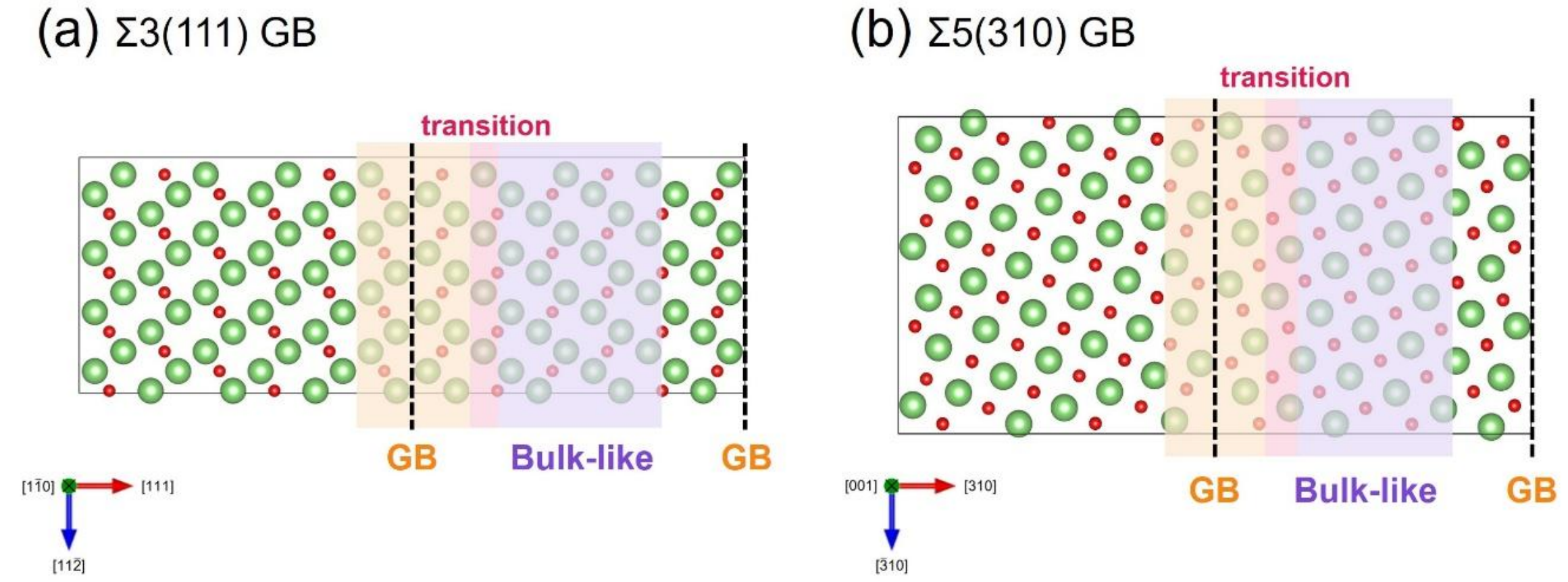

**Figure 2**. Most stable GB structures of Li2O obtained from CNID search and used in this study. (a) ∑3(111) GB and (b) ∑5(310) GB. The black dashed lines indicate the GB planes, and the red arrows indicate the GB normal directions parallel to the applied field directions of the respective GBs. The orange, pink, and purple shaded regions denote the GB, transition, and bulk-like regions, respectively.

In each structure, a Li vacancy was introduced into the bulk-like region, and MD simulations were performed at 600 K under an electric field of 0.100 V/Å applied along the GB normal direction. Detailed simulation conditions are provided in the **Methods** section. Although 600 K is higher than typical battery operating temperatures, this temperature was chosen to ensure sufficient Li-ion hopping within the limited MD timescale.[33] In realistic electrochemical environments, electric fields are spatially nonuniform and substantially stronger local electric fields at the atomic scale have been reported.[34,35] The field strength used in this study was therefore not intended to directly reproduce the macroscopic electric field in an operating battery. Instead, it was selected because orientation-dependent transport in bulk $Li_2O$ becomes more evident at ~0.075 V/Å (**Figure S10**). A field of 0.100 V/Å was chosen to observe field-driven changes more clearly within the accessible MD timescale than at lower fields, while the field-induced reduction in activation energy is quantitatively reproduced (**Supporting Information S1**).

To visualize Li-ion transport, **Figure 3** shows two-dimensional Li occupation probability density maps, together with the corresponding Li relative displacement density maps of representative Li-ion trajectories from three independent MD simulations. The trajectories were projected along the GB normal direction to clearly highlight one GB and its adjacent bulk-like region. The remaining projections are provided in **Figure S11**. Here, enhanced Li-ion mobility refers to migration from the GB region toward the bulk-like region rather than motion within the GB plane. This is because Li-ion transport can extend throughout the cell only after vacancies trapped at the GB are released. Without an electric field, Li-ion mobility in both GBs remained localized near the GB region (**Figure 3**a). The Li vacancy formation energies at the GB were lower than those in the bulk-like region (**Figure S8**). Accordingly, although the Li vacancy was initially introduced into the bulk-like region, it migrated toward the GB and remained there. These results indicate that GBs act as a trapping site for Li vacancies under equilibrium MD conditions.

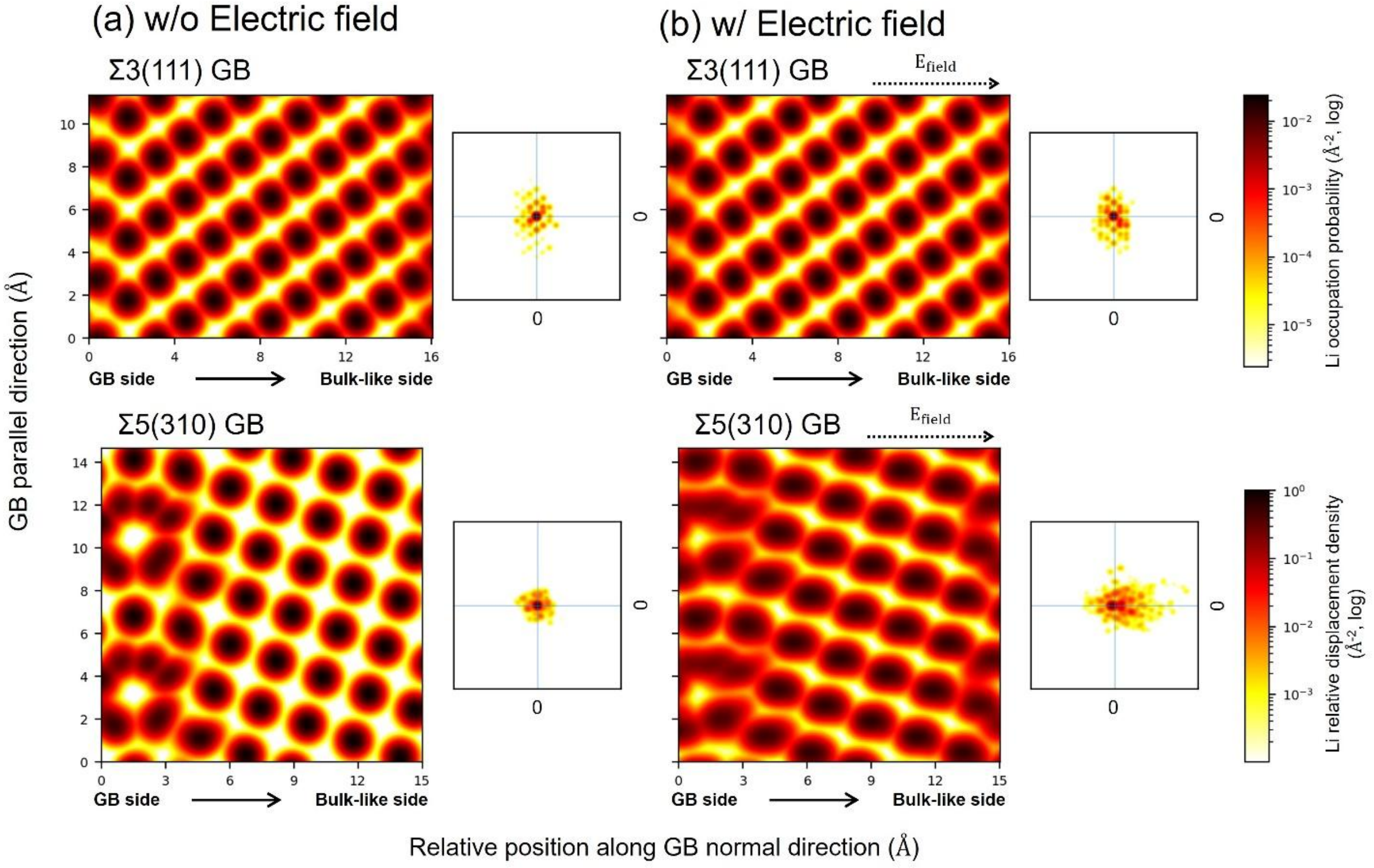

**Figure 3**. Two-dimensional Li occupation probability densities from Li-ion trajectories along the GB normal direction (each left panel) and Li relative displacement densities (each right panel) for representative Σ3(111) and Σ5(310) GBs (a) before and (b) after application of an external electric

field. In each case, brown indicates a higher probability density, whereas yellow indicates a lower probability density. The black arrows indicate the direction of the applied electric field ($E_{field}$). For the left panels, Li positions from the trajectories were projected onto a two-dimensional plane and Gaussian smoothing was applied. The relative position along the GB normal direction runs from the GB side toward the adjacent bulk-like region. For the right panels, the displacement of each Li ion was calculated relative to its initial position and represented as a two-dimensional histogram. Both distributions were normalized so that their integrals equal 1. A broader distribution indicates increased Li-ion mobility along the GB normal direction.

The applied electric field changes this picture in opposite ways for the two GBs (**Figure 3**b). No significant change in the basic GB framework is observed in either structure upon application of the field. In Σ5(310), the vacancy site trapped near the GB disappears, and the occupation probability density between neighboring sites becomes connected along the field direction. The displacement densities also broaden markedly along the field direction, indicating that Li-ion mobility extends from the GB into the bulk-like region. In contrast, Li remains localized at lattice sites in Σ3(111) even under the field, with little increase in mobility along the GB normal direction. The segregation energies calculated by varying the Li vacancy position were approximately 0.4 eV for Σ3(111) and 0.8 eV for Σ5(310) (**Figure S8**). Thus, field-driven Li-ion transport occurs in Σ5(310) despite a segregation energy approximately twice that of Σ3(111), indicating that segregation energy alone does not determine the GB response to an electric field.

To compare the observed Li-ion trajectories, the time-dependent Li-ion mean squared displacement (MSD) along the applied field direction was analyzed separately in the GB and bulk-like regions. As discussed above, the extension of Li-ion mobility into the bulk-like region was used as an important indicator of enhanced Li-ion transport. For quantitative comparison, **R** was defined as the ratio of the mean MSD in the GB region to that in the bulk-like region at 1.0 ns. The **R** value closer to 1 indicates a smaller difference in Li-ion mobility between the GB and bulk-like regions and more spatially extended transport across the cell. The MSD curves and corresponding **R** values are shown in **Figure 4**. In the Σ3(111), the **R** value decreased by approximately 62.0 % after applying the electric field. In the Σ5(310), however, increased Li-ion mobility in the bulk-like region resulted in a 97.1 % decrease in **R**, bringing the value close to 1. A similar trend was also observed not only for these structures but also for the initial GB structures. These results are summarized in **Figure S12**.

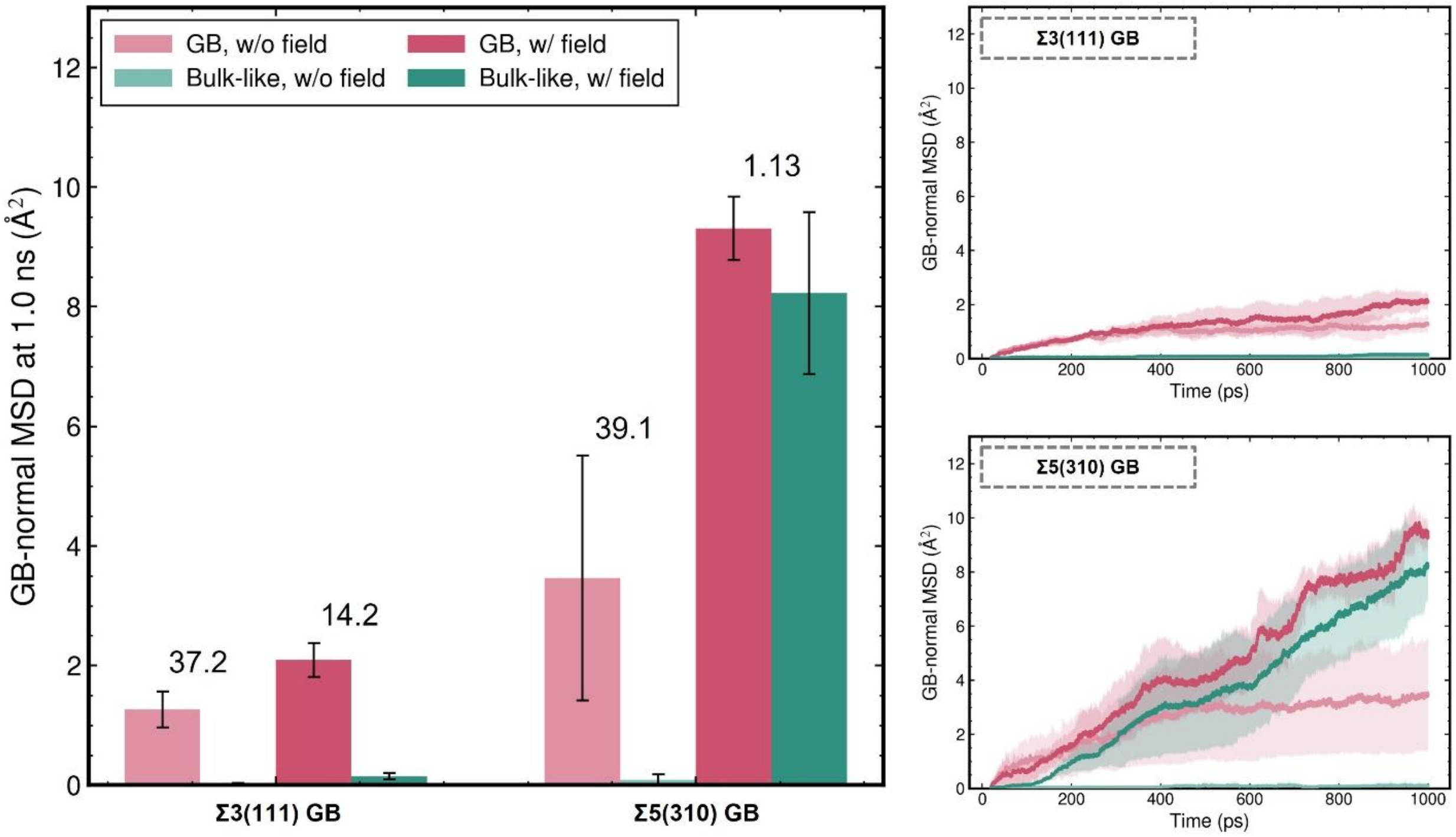

**Figure 4**. Comparison of R values (left) and Li-ion MSDs (right) for the GB structures. R is defined as the ratio of the mean MSD in the GB region to that in bulk-like region at 1.0 ns and is shown above each bar with error bars. Pink and green indicate the GB and bulk-like regions, respectively. Light and dark colors represent the results without and with an electric field applied along the GB normal direction. In the MSD plots, the solid lines represent the mean across three independent simulations, and the shaded regions indicate the standard deviation among them.

Notably, the vacancy is more readily released from the Σ5(310) GB under the electric field even though this GB traps the vacancy more strongly. From a static thermodynamic perspective, the larger segregation energy of Σ5(310) indicates stronger vacancy trapping and thus a lower tendency to release the vacancy. On the other hand, the driving force exerted by the electric field along the GB normal direction is the same for both GBs. Therefore, the opposite responses of the two GBs cannot be explained by either the magnitude of the driving force or the segregation energy. The determining factor is the geometry between the grain orientation and the $Li_2O$ diffusion pathways, specifically the angle between the electric field and the ⟨100⟩ directions available for Li migration. For Σ5, whose GB normal is [310], one of the ⟨100⟩ diffusion pathways has a large projection along the applied electric field, allowing the vacancy to escape from the GB. Indeed, the Li trajectories form straight pathways from the GB into the bulk-like region, inclined by ~18.4° from the field direction, consistent with the 18.4° angle between [310] and [100]. For Σ3, whose GB normal is [111], all six diffusion pathways are equally inclined by 54.7° to the electric field, and the field-induced driving force is insufficient to release the vacancy from the GB. This geometric contrast is also reflected in the field-strength dependence. **Figure S10** shows that the pathways in Σ5(310) begin to appear at 0.050 V/Å and become fully connected at 0.100 V/Å, whereas Σ3(111) exhibits only a weak field response over the same field range.

The approximately 1.8-fold difference among two crystallographic orientations manifests at GBs as a qualitative difference in whether Li-ion transport occurs. This result has two implications. First, GBs that appear to trap vacancies in equilibrium MD can become Li-ion transport pathways under an applied electric field. Therefore, GB evaluation and design should account for the external driving force. Second, field-driven GB transport can be predicted to some extent from the orientations of the constituent grains. Constructing GBs from grains with electric-field responses may provide a design strategy for enhancing Li-ion transport across GBs.

In this study, we developed an electric-field MD framework by combining a fine-tuned MACE potential with a Bader charge prediction model and investigated Li-ion transport in bulk and GB structures of $Li_2O$ under an external electric field. In cubic $Li_2O$, Li-ion diffusion is isotropic without an electric field. However, under the electric field, Li-ion transport strongly depends on crystallographic orientation. The field response is strongest along [100] and [310] orientations, and weakest along [111] orientation. This orientation dependence originates from the restriction of Li diffusion pathways to the ⟨100⟩ direction and can be understood from the angle between the diffusion pathway and the electric field.

At GBs, this orientation dependence leads to a qualitative difference in Li-ion transport. Without an electric field, both Σ3(111) and Σ5(310) GBs trap Li vacancies. The segregation energy of Σ5(310) is approximately twice that of Σ3(111), indicating stronger vacancy trapping at the Σ5(310) from a static thermodynamic perspective. Under an external electric field, however, the vacancy escapes from Σ5(310) and Li-ion transport extends from the GB into the bulk-like region. In contrast, the vacancy remains at the Σ3(111) despite its weaker trapping tendency. These results demonstrate that GB transport under the electric field is governed by the crystallographic orientations of the constituent grains rather than solely by static thermodynamic quantities such as segregation energy.

GBs that appear to act as trapping sites in equilibrium MD can become conduction pathways under an electric field. This finding provides a new perspective for designing GBs in SEIs and solid electrolytes based on their function under an external driving force rather than their stability at equilibrium.

## Computational Method

### Dataset Construction

To construct the dataset for the fine-tuned MACE potential and charge prediction model of $Li_2O$, the on-the-fly machine learning force field implemented in the Vienna Ab initio Simulation Package (VASP) was used.[36–39] During MD simulations, the method continuously estimates the prediction uncertainty of the machine learning force field. When the Bayesian error in the forces exceeds a predefined threshold, first-principles calculations are performed for the corresponding atomic configuration.

The initial structure was a 2×2×2 supercell of cubic $Li_2O$ with the $Fm\bar{3}m$ structure. To obtain diverse local Li environments, structures containing different numbers and arrangements of Li vacancies were used. The cell length was 9.27 Å and Li vacancy concentrations of up to 18.75 % were considered. Structural optimization was performed using VASP, the Perdew–Burke–Ernzerhof (PBE) functional within the generalized gradient approximation (GGA) was used for the exchange-correlation energy and electron–ion interactions were described using the projector augmented-wave (PAW) method.[40,41] A plane-wave cutoff energy of 750 eV and a Γ-centered 4×4×4 k-point mesh was used. Atomic positions were optimized until the maximum force on each atom was less than 0.01 eV/Å.

On-the-fly learning was performed in the NpT ensemble over a temperature range of 1 ~ 1000 K with a time step of 1.0 fs. The first-principles reference calculations used the same settings as the structural optimizations except for Brillouin-zone sampling, for which a Γ-centered 2×2×2 k-point mesh was used. This process generated a total of 14,988 bulk configurations.

To supplement local coordination environments around GBs, Σ5(100) and Σ5(310) GB structures were additionally included. For each GB, pristine structures and defective structures containing Li vacancies at different positions and concentrations up to 2.5 % were considered. Structural optimizations were performed using Γ-centered 5×5×2 and 6×4×1 k-point meshes, respectively. This generated 51 GB configurations, resulting in a final dataset of 15,039 structures.

**Machine-learned model development**

The constructed dataset was used to fine-tune the pretrained universal MLIP MACE-MP-0 large model and develop an MLIP specialized for $Li_2O$. [42,43] The dataset was divided into 80 % for training, 10 % for validation, and 10 % for test. GB configurations were distributed across all three sets. The model was trained with a batch size of 4 and an initial learning rate of 0.01. Equal weights were assigned to the energy and force terms in the loss function. The maximum number of training epochs was set to 100, with an early stopping patience of 20 epochs. Exponential moving average (EMA) and stochastic weigh averaging (SWA) were used to stabilize the training, and AMSGrad was employed as the optimizer.

**Bader charge calculation and charge prediction model**

To generate the Bader charge dataset used as the training target for the charge prediction model, first-principles DFT calculations were performed using VASP with the constructed dataset as reference structures.[37,44,45] The PBE functional within the GGA was used for the exchange-correlation energy, and electron-ion interactions were described using PAW method.[40,41] A Γ-centered 2×2×2 k-point mesh and a plane-wave cutoff energy of 750 eV were used. Electronic self-consistent calculations were converged to an energy criterion of $10^{-6}$ eV. Bader charges were obtained by partitioning the valence charge density, using the sum of the core and valence electron densities as the reference density.[23]

A DimeNet++ model was constructed to predict structure-dependent Bader charges in complex local atomic environments including GBs.[46,47] DimeNet++ uses atomic numbers and atomic coordinates as inputs. It represents local atomic environments through directional message passing that incorporates interatomic distances and angular information from neighboring atomic triplets.

Atomic graphs were constructed using the ASE neighbor list with periodic boundary conditions, and the interaction cutoff was set to 5.0 Å. The model was modified to directly output one Bader charge for each atom. The same training, validation, and test structures used for MACE fine-tuning were employed, with Bader charges as target values. The model used 128-dimensional hidden representations and four interaction blocks. A batch size of 16 and an initial learning rate of 0.0001 were used.

**Molecular Dynamics with Electric Field**

MD simulations were performed using the Atomic Simulation Environment (ASE) with a Langevin thermostat in the NVT ensemble at 600 K.[48] Initial velocities were generated according to the Maxwell–Boltzmann distribution. A time step of 1 fs was used, and each simulation was performed for $1.0 \times 10^6$ MD steps corresponding to 1.0 ns. Unless otherwise stated, an electric field of 0.100 V/Å ($10^7$ V/cm) was applied; for the [100], [310], and [111] bulk structures and the GBs, 0.050 and 0.075 V/Å were additionally examined (**Figures S6 and S10**). To obtain a measurable drift that is not obscured by thermal diffusion within the nanosecond timescale accessible to MD simulations, the work done by the electric field during a single Li hop must be comparable to or greater than the thermal energy.[5] At 0.100 V/Å, this work is ~0.2 eV, about four times the thermal energy at 600 K (0.052 eV) and close to the zero-field activation barrier of 0.244 eV. Therefore, the applied field represents a strong-field regime relevant to local fields near interfaces.

A uniform external electric field was applied along the specified c axis for the bulk structures and along the GB normal direction for the GB structures. The first 20 ps of each trajectory were excluded from the Li-ion trajectory analysis to eliminate the initial equilibration period.

To account for the force induced by the external electric field, the structure-dependent scalar Bader charge $q_i^{\text{Bader}}$ of each atom *i* was predicted using the DimeNet++ charge prediction model. The predicted charges were represented in the following diagonal matrix:

$$Q_i^{\text{Bader}} = q_i^{\text{Bader}}\mathbf{I} = \begin{pmatrix} q_i^{\text{Bader}} & 0 & 0 \\ 0 & q_i^{\text{Bader}} & 0 \\ 0 & 0 & q_i^{\text{Bader}} \end{pmatrix} \tag{1}$$

Accordingly, the additional force acting on atom *i* under an electric field $E_{\text{field}}$ is given by:

$$F_i^{\text{Bader}} = Q_i^{\text{Bader}} E_{field} \tag{2}$$

Using this expression, the total force acting on atom $i$ in the MD simulations was defined as the sum of the force predicted by the fine-tuned MACE potential and the force induced by the external electric field.

$$F_i = F_i^{\mathrm{MACE}} + F_i^{\mathrm{Bader}} \quad (3)$$

## AUTHOR INFORMATION

### Corresponding Author

Teruyasu Mizoguchi — *Institute of Industrial Science, The University of Tokyo, Tokyo 153-8505, Japan* and *Department of Materials Engineering, The University of Tokyo, Tokyo 113-8656, Japan*; https://orcid.org/0000-0003-3712-7307; Email: teru@iis.u-tokyo.ac.jp

### Authors

Youjeong Choi — *Department of Materials Engineering, The University of Tokyo, Tokyo 113-8656, Japan*

Po-Yen Chen — *Department of Materials Engineering, The University of Tokyo, Tokyo 113-8656, Japan*

Tasuku Sugiura — *Institute of Industrial Science, The University of Tokyo, Tokyo 153-8505, Japan*

### Funding Sources

This work was supported by the Japan Science and Technology Agency (JPMJGX24S0).

### Notes

The authors declare no competing financial interest.

## ACKNOWLEDGMENT

This work was supported by the Japan Science and Technology Agency (JST) as part of Green Technologies of Excellence (GteX), Grant Number JPMJGX24S0. The computation was carried out using the computer resource offered by Research Institute for Information Technology, Kyushu University.

## ASSOCIATED CONTENT

**Supporting Information Available:**

The supporting information is available free of charge on the ACS Publications website.

Model validation results, bulk orientations and GB structures, Li vacancy formation energies, electric field strength dependence and additional figures (PDF)

# Supporting Information

## Electric Fields Turn Vacancy-trapping Grain Boundaries into Li-Ion Pathways in $Li_2O$

Youjeong Choi [a], Po-Yen Chen [a], Tasuku Sugiura [b], and Teruyasu Mizoguchi [a, b]*

[a] Department of Materials Engineering, The University of Tokyo, Tokyo 113-8656, Japan
[b] Institute of Industrial Science, The University of Tokyo, Tokyo 153-8505, Japan
* Email: teru@iis.u-tokyo.ac.jp

## S1. Model validation

**Figure S1a** shows the predictive accuracy of the resulting fine-tuned MACE potential for total energies and atomic forces evaluated using an independent test set of 1,503 configurations. The model achieved mean absolute errors (MAEs) of 4.43 meV/atom for energy and 13.86 meV/Å for atomic force. It demonstrates that the fine-tuned MACE potential accurately reproduces DFT-calculated energies and forces across the structural space considered in this study. We further evaluated the ability of the model to describe the phonon properties of $Li_2O$, as shown in **Figure S2**.[1,2] The phonon dispersion and phonon density of states of pristine $Li_2O$ calculated using the fine-tuned MACE potential show good agreement with the corresponding DFT results. Good agreement with DFT was also obtained for structure containing Li vacancy, indicating that the model can accurately describe phonon properties not only in pristine $Li_2O$ but also in local defect environments introduced by Li vacancy.

To approximate the forces acting on atoms under an electric field, we first attempted to construct a GNN model based on Born effective charges (BECs). To obtain BECs for diverse atomic environments, we constructed three datasets consisting of defect-free structures, structures containing Li vacancies at various concentrations, and structures containing electrically neutral Schottky defects formed by removing two Li ions and one O ion. The first two datasets were constructed using a subset of the structures generated for the Bader charge calculations. For the Schottky defect dataset, ten initial structures were generated by removing Li–Li–O combinations with different spatial arrangements from $Li_{64}O_{32}$. These structures were then subjected to structural optimization and on-the-fly learning using the same procedure as that employed to generate the structures for the Bader charge dataset. Reference BECs were calculated using density functional perturbation theory (DFPT) as implemented in VASP.[3] The computational settings were consistent with those used for the Bader charge calculations, with a Γ-centered 2×2×2 k-point mesh for Brillouin-zone sampling and an energy convergence criterion of $10^{-6}$ eV for the electronic self-consistent calculations. Considering the computational cost of BEC calculations, a convergence test was performed for plane-wave cutoff energies from 520 to 750 eV. The root mean square (RMS) difference in BEC values at 520 eV relative to 750 eV was $4.23 \times 10^{-4}$ e. Based on the balance between computational cost and accuracy, a cutoff energy of 520 eV was used.

We constructed a BEC prediction model using the Equival_evar model developed by Kutana et al.[4] The pretrained BM1 model of Equivar_eval was fine-tuned separately using defect-free and defective structures. For the defect-free dataset, 820 pristine structures were divided into 524 training, 132 validation, and 164 test constructions. For the defective dataset, 3,037 structures were divided into 2,193 training, 244 validation, and 600 test structures. For the schottky dataset, 1,000 structures were divided into 810 training, 90 validation, and 100 test structures. A batch size of 4 and an initial learning rate of 0.0001 were used. Early stopping was applied with a patience of 20 epochs.

The MAE validation results for the fine-tuned Equivar_eval models trained on the three datasets are shown in **Figure S3(a, b, c)**. For the defect-free dataset, the MAEs across all BEC components relative to the DFPT results were 0.0027 e for Li and 0.0033 e for O. In contrast, different behaviors were observed for the defective structures. For the dataset containing Li vacancies, similar local atomic environments produced substantially different BEC values as shown in **Figure S3d**. Anomalous BEC values were also obtained from the DFPT calculations, consistent with issues reported in previous studies.[4,5] As a result, the model showed relatively large MAEs of 0.4925 e for Li and 0.6261 e for O. The model trained on structures containing Schottky defects also showed relatively larger errors than the defect-free model, with MAEs of 0.0158 e for Li and 0.0212 e for O. These results indicate that obtaining stable and consistently predictable BEC values becomes more difficult in defective structures. Such low prediction accuracy for defective structures can directly affect the field-induced forces and may reduce the reliability of charge and force predictions during MLIP-MD simulations.

Based on these results, we constructed a Dimenet++ based Bader charge prediction model using Bader charges as training targets for structure-dependent effective partial atomic charge.[6,7] The charge prediction accuracy of the model is shown in **Figure S1b**. The MAEs relative to the DFT results were 0.0005 e for Li and 0.0050 e for O. The MAE of the total Bader charge was 0.0905 e per structure. These results confirm that the predicted atomic Bader charges reproduce the DFT based values with high accuracy.

In addition, electric-field MLIP-MD simulations were performed for perfect $Li_2O$ (324 atoms) to compare the Bader charge prediction model with BEC prediction model trained on defect-free $Li_2O$ structures. A field of 0.10 V/Å was applied along the c axis at 600 K, and three independent 1.0 ns simulations were performed for each model. The mean Li MSDs obtained from the two models are shown in **Figure S4**. The difference between the mean Li MSD values was approximately 3.6 % confirming similar Li-ion dynamics during the MD simulations.

To perform MD simulations under an applied electric field, we constructed an electric-field MLIP-MD framework by combining the fine-tuned MACE potential with the Dimenet++ Bader charge prediction model. **Figure S1c** shows the calculated activation energies for bulk $Li_2O$ containing a Li vacancy, which were used to evaluate the ability of this framework to describe Li-ion migration. MD simulations were performed for 0.50 ns at 300, 400, 500, and 600 K under electric fields of 0, 0.050, 0.075, 0.100 V/Å. The activation energy was determined from the temperature dependence of the Li-ion diffusion coefficient under each condition.

When an effective charge ($q$) moves a distance ($\Delta Z$) along a constant electric field ($E$), the work performed by the electric field is expressed as $qE \cdot \Delta Z$. We estimated the range of the field-induced reduction in the Li-ion hopping activation energy $W_{field}$, using the hopping distance ($Z_{hop}$). The lower bound assumes that the transition state is located halfway along the hopping distance and the upper bound corresponds to the work performed over the full hopping distance. [8,9] The resulting expressions are given below.

$$qE \cdot \frac{1}{2} Z_{hop} \leq W_{field} \leq qE \cdot Z_{hop}$$

Without an electric field, the activation energy was 0.24 eV. This value is close to the reported migration barrier range of approximately 0.15 ~ 0.33 eV for vacancy-mediated Li diffusion.[10,11] The activation energy reductions at 0.075 and 0.100 V/Å agreed well with the estimated range, whereas a deviation was observed at 0.050 V/Å. These results indicate that the proposed model can reasonably capture the trend of field-driven Li-ion migration in the relatively high electric field regime.

The Σ3(111) GB used in the subsequent GB calculations was not included in the training data of the fine-tuned MACE potential. To evaluate its transferability to this GB, the GB energy and the site-dependent Li-vacancy segregation energy were recalculated by DFT using the same settings as those used for the dataset construction. The GB energy was 0.663 J/m$^2$ with the fine-tuned MACE potential and 0.659 J/m$^2$ with DFT. The corresponding Li-vacancy segregation energies were 0.421 eV and 0.450 eV, respectively. Thus, the differences between the fine-tuned MACE potential and DFT were less than 0.005 J/m$^2$ for the GB energy and 0.03 eV for the segregation energy.

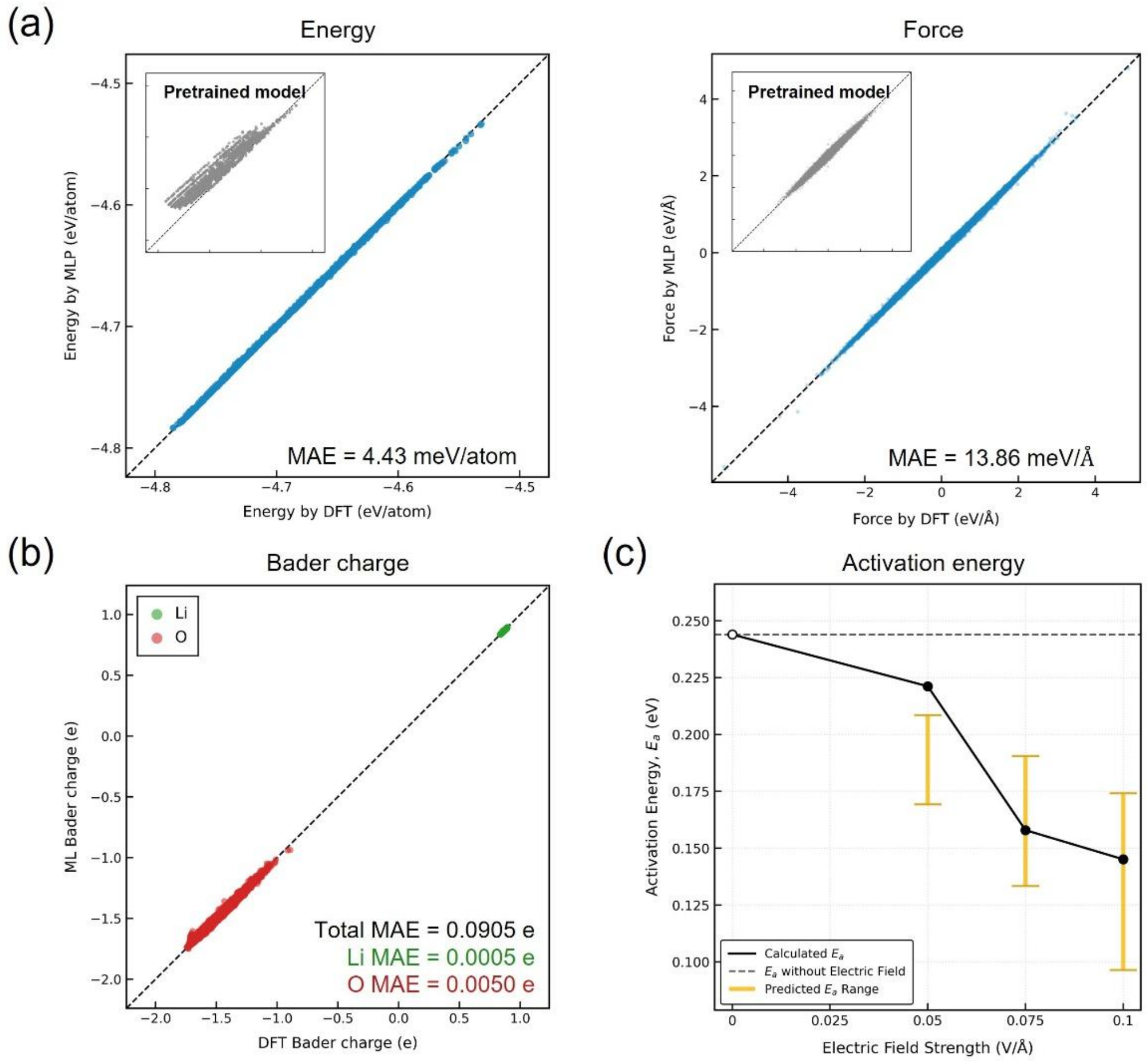

**Figure S1**. (a) Comparison of energy (left) and atomic force (right) calculated using DFT and the fine-tuned MACE potential. The insets show the corresponding results from the pretrained MACE-MP-0. (b) Comparison of the Bader charges of Li and O atoms obtained from DFT and the developed Bader charge prediction model. (c) Activation energies for Li-ion diffusion in bulk $Li_2O$ under external electric fields and the predicted ranges of field-induced activation energy reduction.

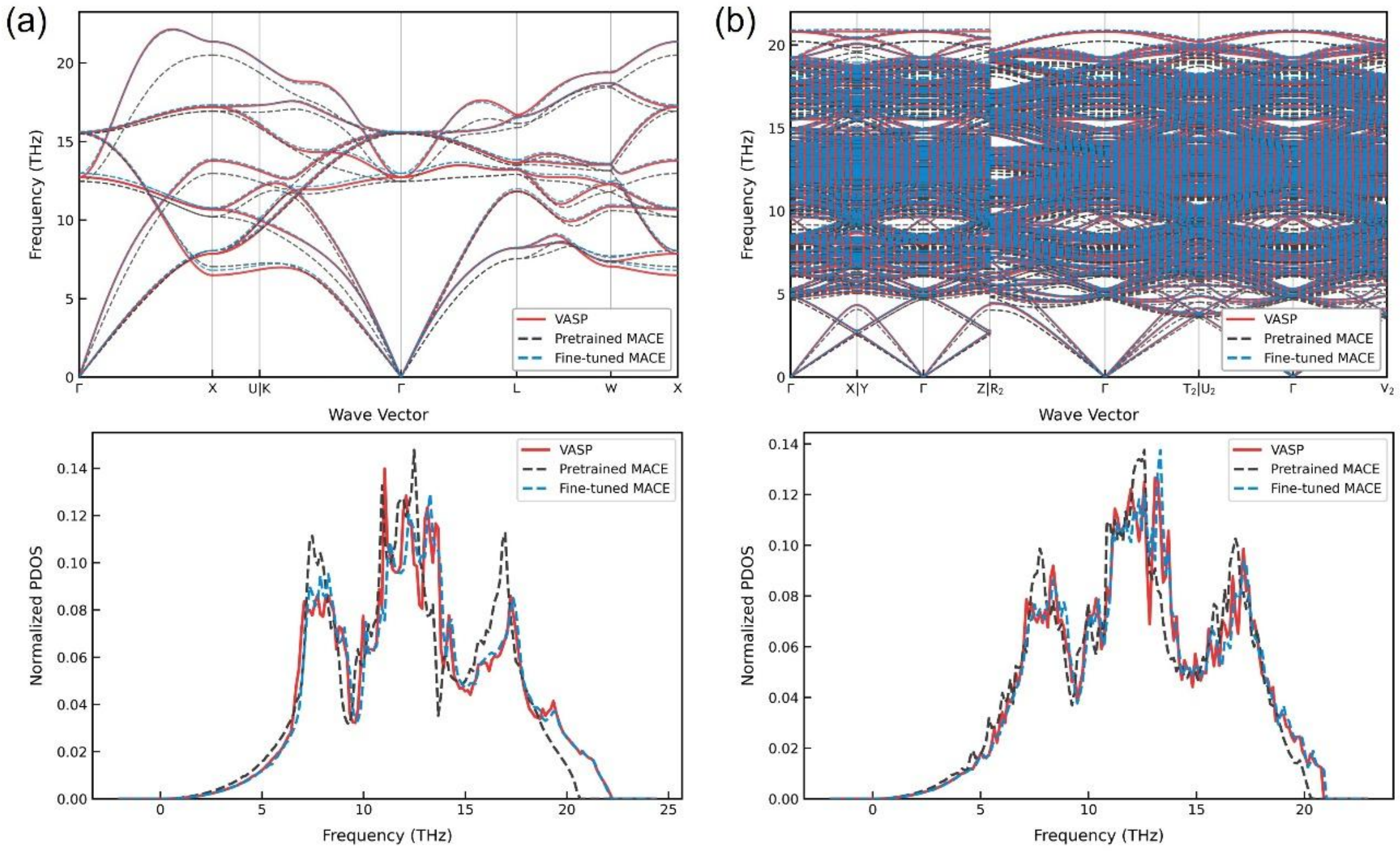

**Figure S2**. Comparison of the phonon dispersion and phonon density of states (PDOS) of the $Li_2O$ calculated using DFT, the pretrained MACE, the fine-tuned MACE potential. (a) Pristine $Li_{64}O_{32}$ (b) $Li_{63}O_{32}$ with Li vacancy.

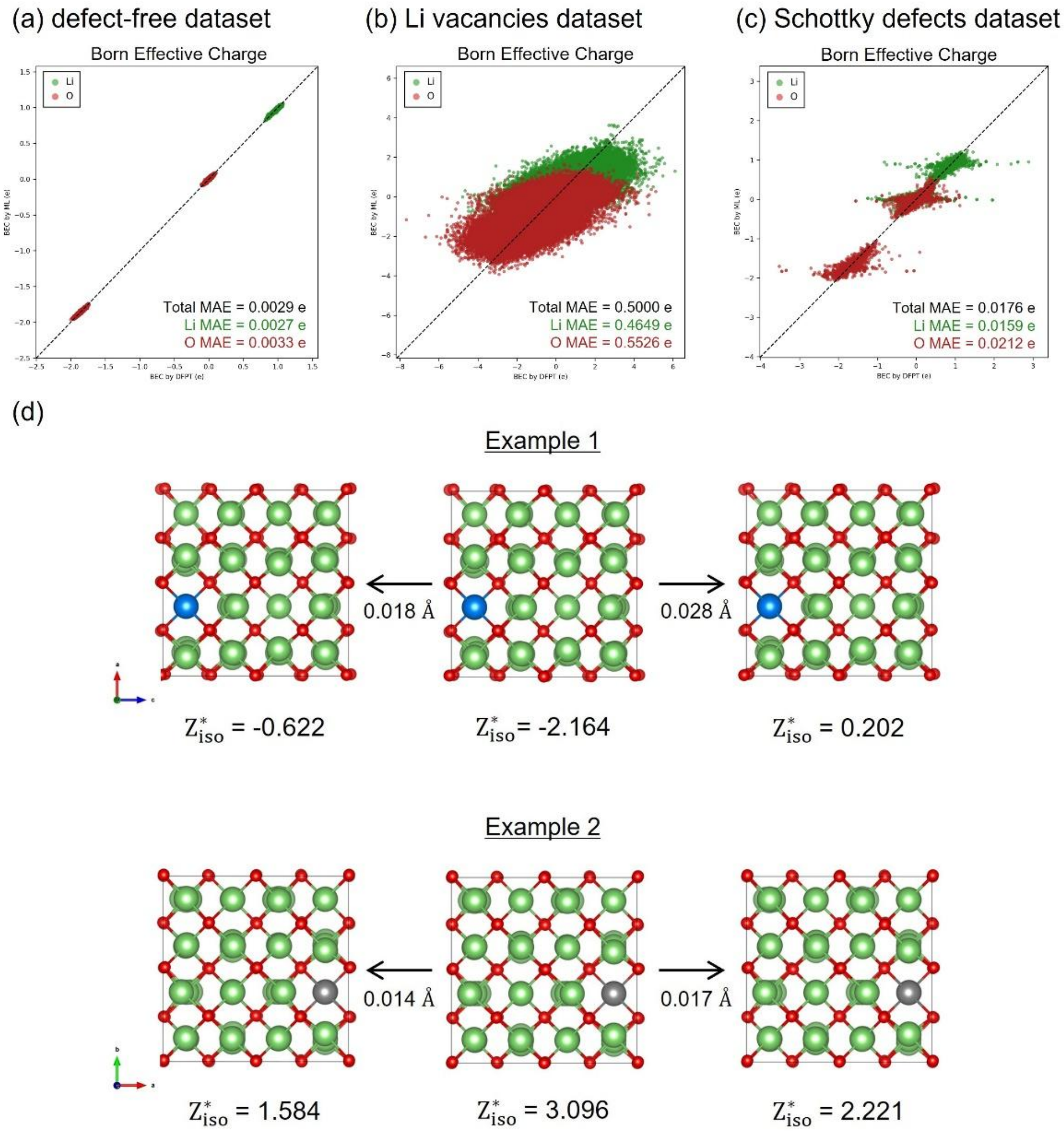

**Figure S3**. MAEs of all BEC components for Li and O atoms obtained with each BEC prediction model. Validation results for models trained using (a) defect-free dataset (b) dataset containing Li vacancies, and (c) dataset containing Schottky defects. Green and red color indicate the results for Li and O, respectively. (d) Examples of structures with similar atomic environments but markedly different BEC values. Blue and gray color indicate the target Li atoms for which the BEC values were evaluated. Using the central structure as reference, the arrows indicate the RMSD of the atomic coordinates around the target Li atom. The BEC values are given as the average of the diagonal components.

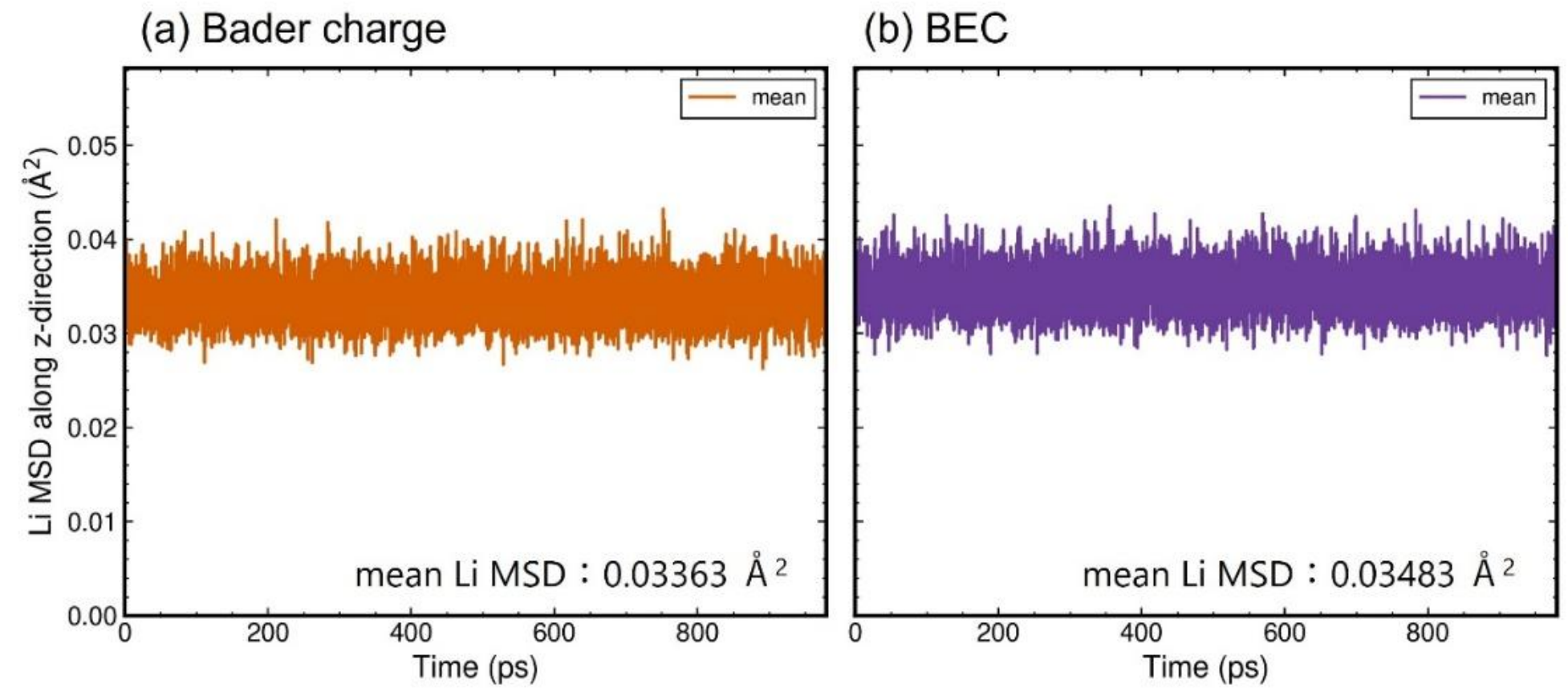

**Figure S4**. Comparison of the mean Li MSDs in perfect $Li_2O$ obtained using (a) Bader charge prediction model and (b) BEC prediction model trained on defect-free dataset.

## S2. Electric-Field Response of bulk $Li_2O$ along different crystallographic orientations

**Table S1**. Cell composition, Li vacancy concentration $c_v$, and cell length along the applied electric-field direction for bulk $Li_2O$ models with different crystallographic orientations.

| Orientation | composition | Li vacancy concentration | Cell length along electric-field direction [Å] |
|---|---|---|---|
| 100 | $Li_{215}O_{108}$ | 0.0046 | 13.90 |
| 110 | $Li_{255}O_{128}$ | 0.0039 | 13.11 |
| 111 | $Li_{287}O_{144}$ | 0.0035 | 16.05 |
| 210 | $Li_{319}O_{160}$ | 0.0031 | 14.66 |
| 211 | $Li_{287}O_{144}$ | 0.0035 | 13.90 |
| 221 | $Li_{215}O_{108}$ | 0.0046 | 13.90 |
| 310 | $Li_{319}O_{160}$ | 0.0031 | 14.66 |
| 311 | $Li_{263}O_{132}$ | 0.0038 | 15.37 |
| 320 | $Li_{311}O_{156}$ | 0.0032 | 16.71 |
| 321 | $Li_{223}O_{112}$ | 0.0045 | 17.34 |

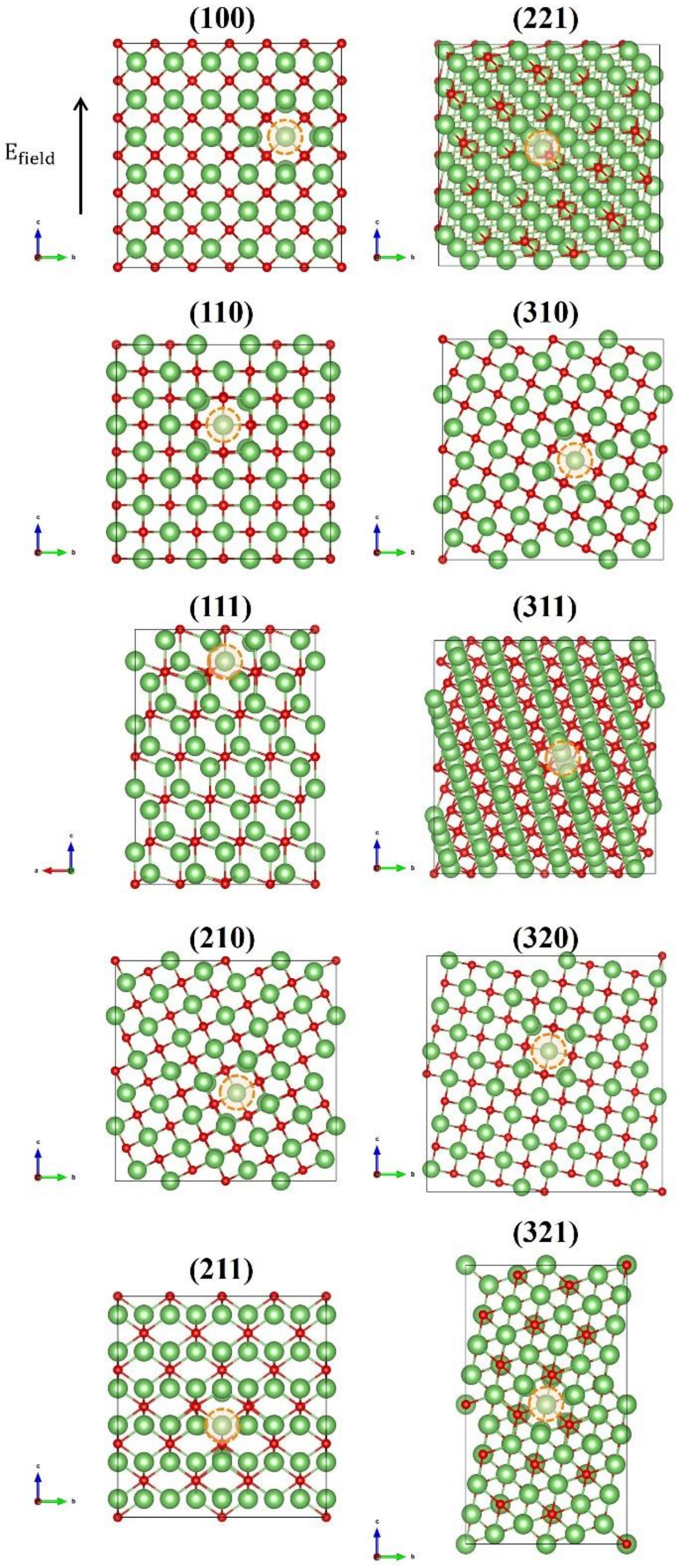

**Figure S5**. $Li_2O$ structures for the ten crystallographic orientations considered in this study. The applied electric field was directed along the c axis in all cases, as indicated by the blue arrows. The orange circles indicate each Li vacancy.

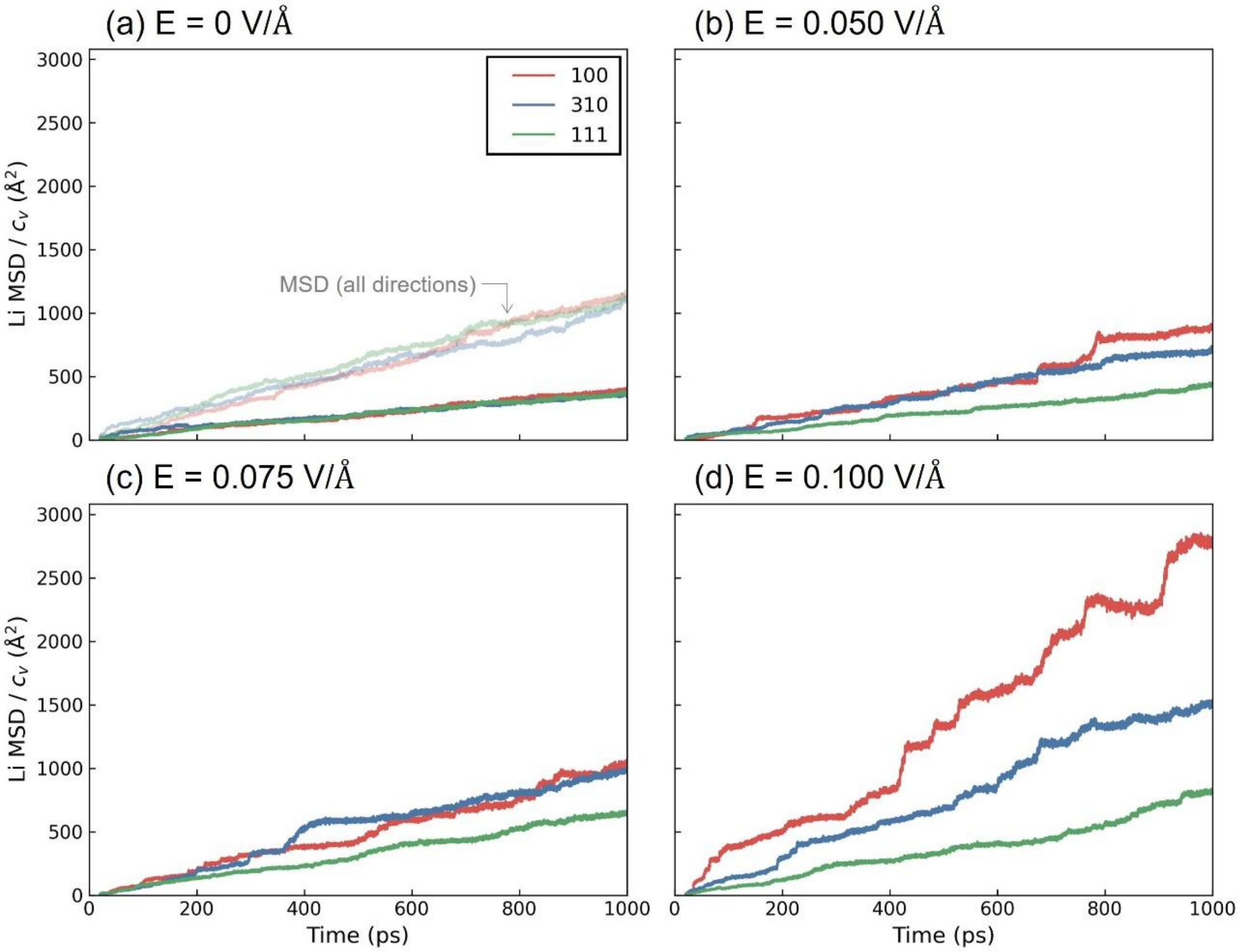

**Figure S6**. Li ion MSD normalized by the Li vacancy concentration for the [100], [310], and [111] orientations under electric fields of (a) 0, (b) 0.050, (c) 0.075, and (d) 0.100 V/Å. The solid lines show the MSD along the field direction. In (a) E = 0 V/Å, the MSD along the same direction is shown for direct comparison with the field-applied results, while the lighter dashed lines indicate the MSD over all directions because Li diffusion at zero field is essentially isotropic. Each curve represents a single 1.0 ns MD trajectory.

## S3. Li-ion transport in $Li_2O$ grain boundaries

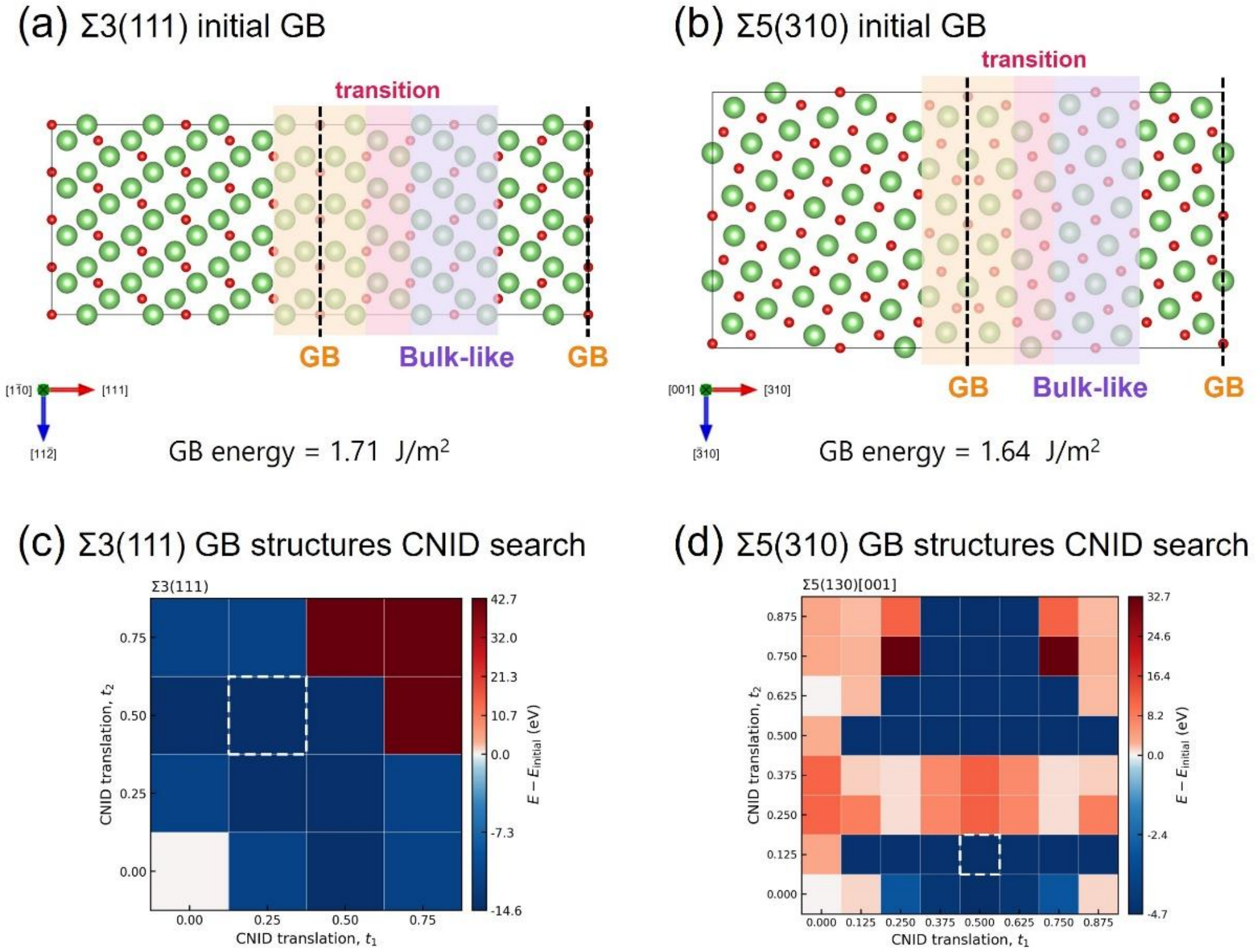

**Figure S7**. $Li_2O$ grain boundary structures with GB energies. (a) Σ3(111) initial GB (b) Σ5(310) initial GB. The black dashed lines indicate the GB planes, and the red arrows indicate the GB normal directions parallel to the applied electric-field directions of the respective GBs. The orange, pink, and purple shaded regions denote the GB, transition, and bulk-like regions, respectively. (c) and (d) show the CNID search results for the respective GB structures. Energy comparisons were performed using a 4×4 CNID translation grid in (c) and an 8×8 grid in (d). The relative energy was calculated with respect to the mirror-symmetric initial structure as $\Delta E = E - E_{\text{initial}}$. $t_1$ and $t_2$ denote fractional translations along the CNID basis vectors, and the white dashed boxes indicate the CNID GB structures with the lowest energies.

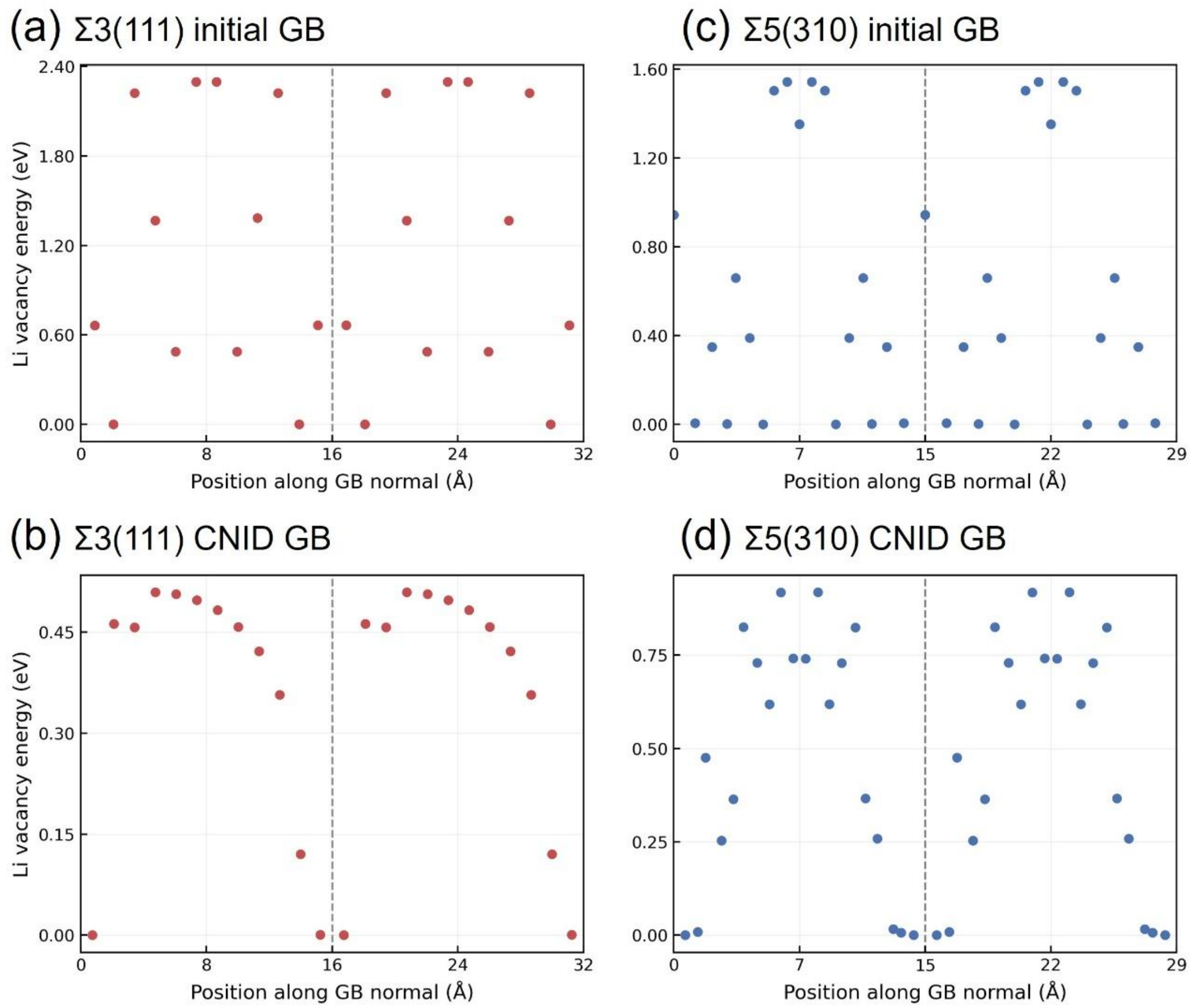

**Figure S8**. Distribution of the Li vacancy formation energy along the GB normal direction. (a) initial and (b) CNID Σ3(111) GBs, (c) initial and (d) CNID Σ5(310) GBs.

The GBs correspond to the low-energy structures used in the main text. To examine the relative variation within each structure while neglecting the effect of the Li chemical potential, the lowest energy in each structure was set to 0.00 and the energy differences relative to that value are shown. The black dashed line indicates the central GB, and lower Li vacancy formation energies are observed near the GB region.

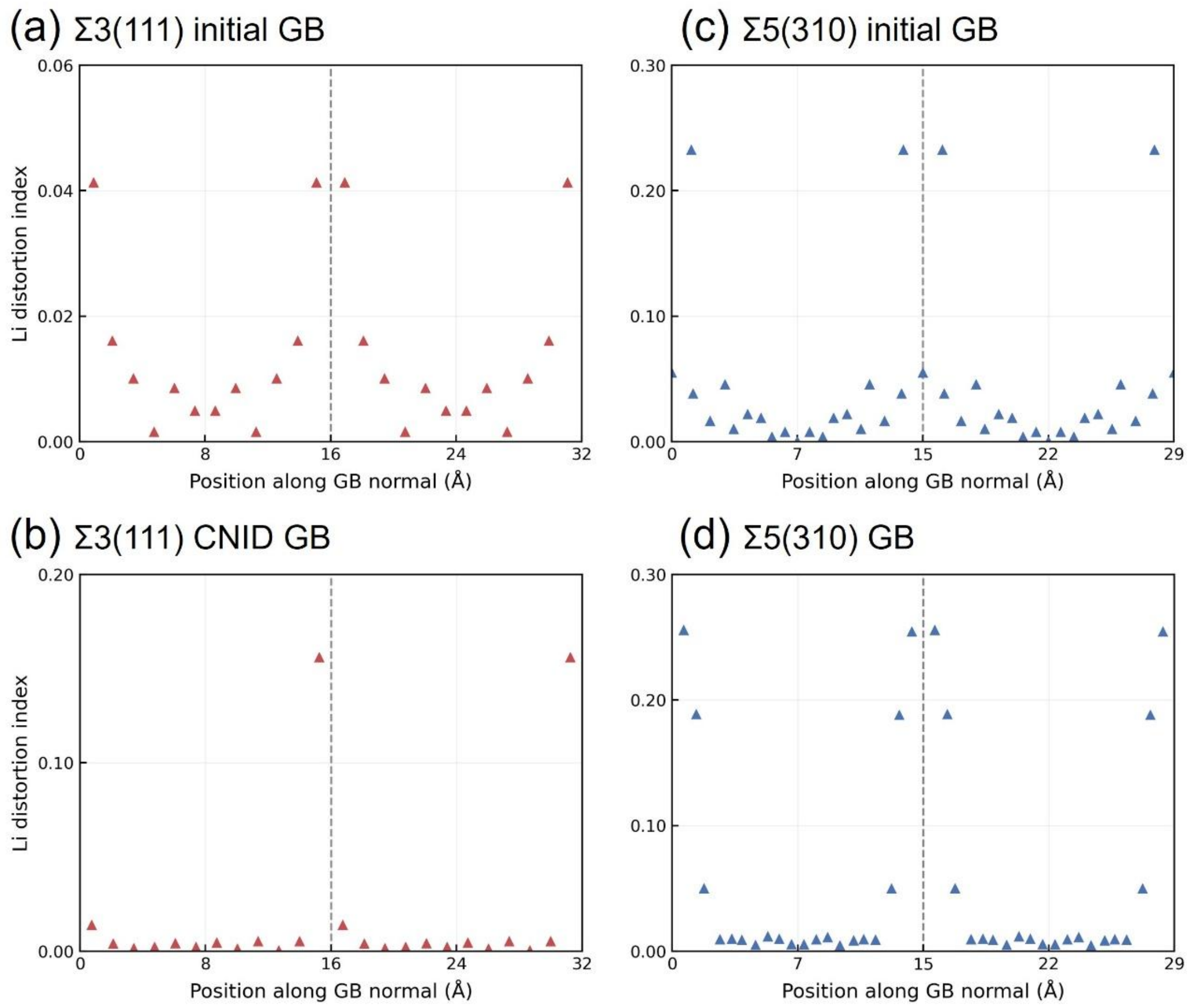

**Figure S9**. Distribution of the local distortion index of Li along the GB normal direction. (a) initial and (b) CNID Σ3(111) GBs, (c) initial and (d) CNID Σ5(310) GBs. The CNID GBs correspond to the low-energy structures used in the main text. The local distortion index was calculated for each Li atom from the relative deviation of the Li–O bond lengths to its four nearest O atoms, with the minimum value set to 0.00. The black dashed line indicates the central GB. Larger distortions are observed in the vicinity of the GB region.

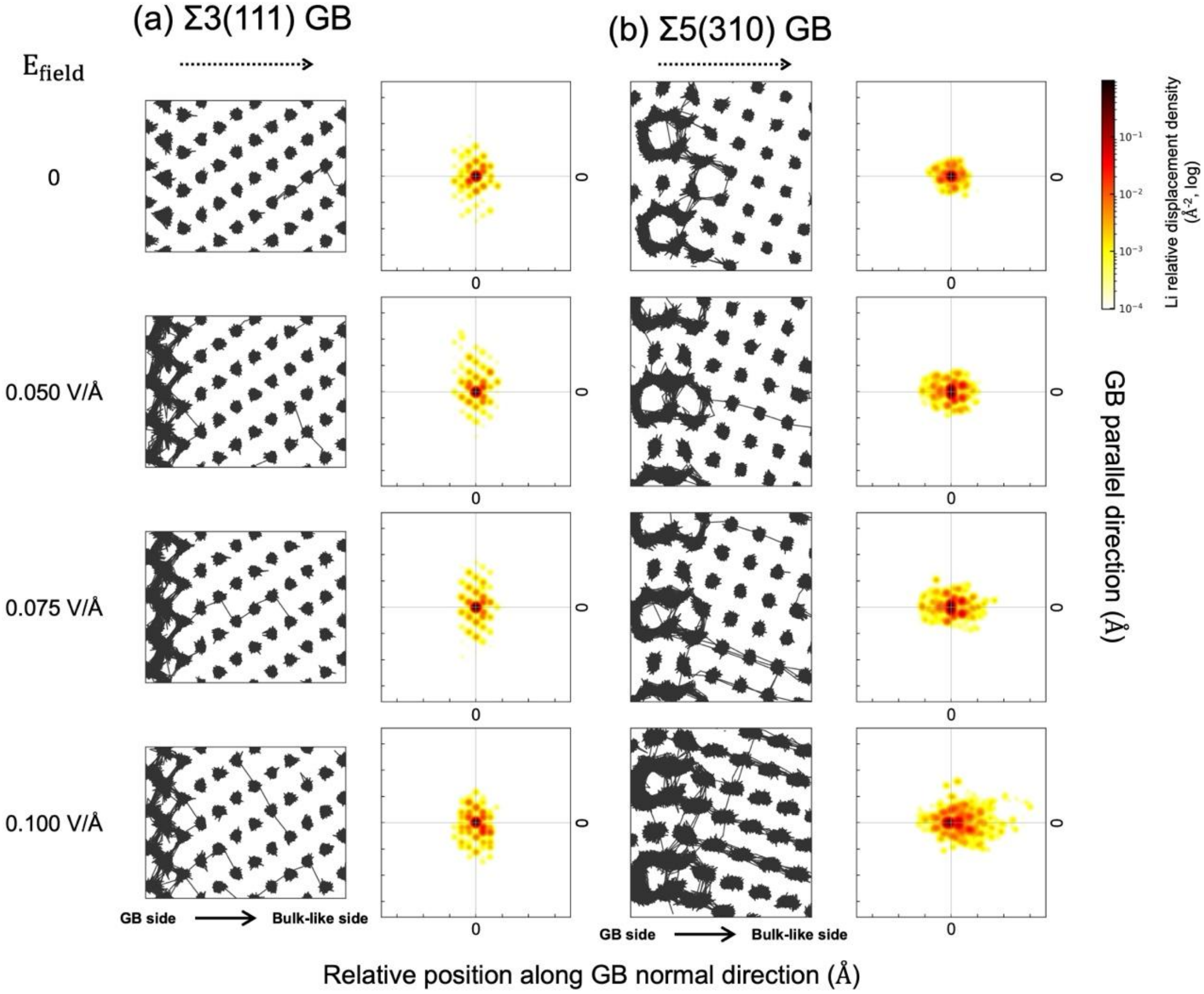

**Figure S10**. Changes in Li-ion mobility with electric field strength for (a) Σ3(111) and Σ5(310) GBs. The rows from top to bottom correspond to electric field strengths of 0, 0.050, 0.075, and 0.100 V/Å, respectively. In the left panels, Li-ion positions obtained from the MD trajectory were projected onto a two-dimensional plane, and the migration paths of individual Li ions were represented by connecting their positions over time. The relative position along the GB normal direction extends from the GB toward the adjacent bulk-like region. In the right panels, the displacement of each Li ion was calculated relative to its initial position and represented as a two-dimensional histogram. The distributions were normalized so that their integrals equal 1. A broader distribution along the horizontal direction indicates increased Li-ion mobility along the GB normal direction.

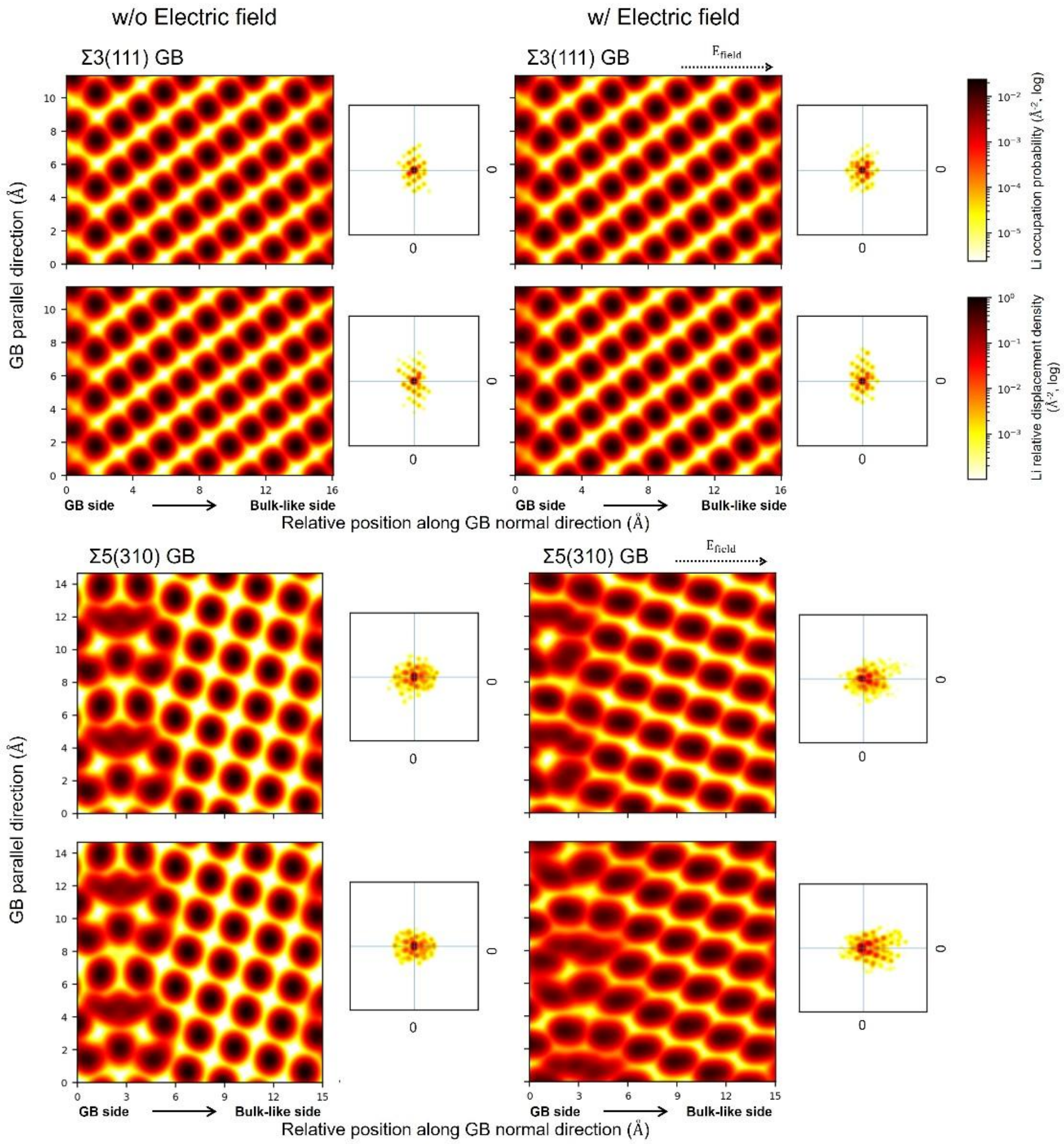

**Figure S11**. Two-dimensional Li occupation probability densities from Li-ion trajectories along the GB normal direction (each left panel) and Li relative displacement densities (each right panel) for remaining Σ3(111) and Σ5(310) GBs (a) before and (b) after application of an external electric field. In each case, brown indicates a higher probability density, whereas yellow indicates a lower probability density. The black arrows indicate the direction of the applied electric field ($E_{field}$). For the left panels, Li positions from the trajectories were projected onto a two-dimensional plane and Gaussian smoothing was applied. The relative position along the GB normal direction runs from the GB side toward the adjacent bulk-like region. For the right panels, the displacement of each Li ion was calculated relative to its initial position and represented as a two-dimensional histogram. Both distributions were normalized so that their integrals equal 1. A broader distribution indicates increased Li motion along the GB normal direction.

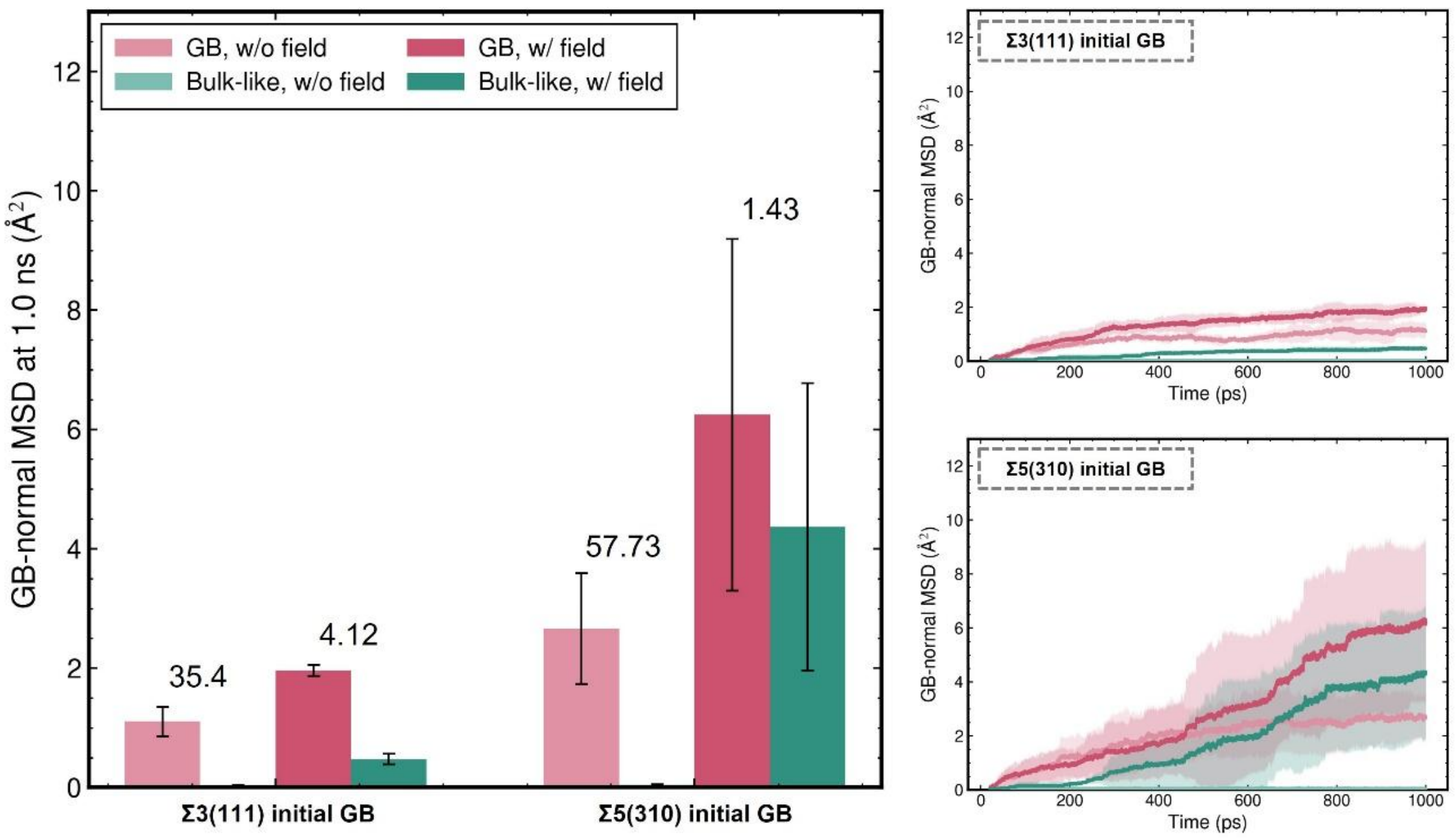

**Figure S12**. Comparison of **R** values (left) and Li-ion MSDs (right) for the Σ3(111) initial GB and Σ5(310) initial GB structures. **R** is defined as the ratio of the mean MSD in the GB region to that in bulk-like region at 1ns and is shown above each bar with error bars. Pink and green indicate the GB and bulk-like regions, respectively. Light and dark colors represent the results without and with an electric field applied along the GB normal direction. In the MSD plots, the solid lines represent the mean of three independent simulations, and the shaded regions indicate the standard deviation among them.